\documentclass[twocolumn,showpacs,preprintnumbers,amsmath,amssymb]{revtex4}
\usepackage[dvips]{graphics}
\usepackage{color}

\newcommand{\beq}{\begin{equation}}
\newcommand{\eeq}{\end{equation}}
\newcommand{\bma}{\begin{math}}
\newcommand{\ema}{\end{math}}
\newcommand{\beqa}{\begin{eqnarray}}
\newcommand{\eeqa}{\end{eqnarray}}
\def\opone{\le\textbf{}\textbf{}avevmode\hbox{\small1\kern-3.8pt\normalsize1}}

\newcommand{\be}[1]{     \begin{eqnarray} \mbox{$\label{#1}$}   }

\newcommand{\ee}{\end{eqnarray}}
\newcommand{\pref}[1]{(\ref{#1})}

\newcommand\ie {{\it i.e. }}
\newcommand\eg {{\it e.g. }}

\newcommand{\ul}\underline

\begin{document}

\title{The Pfaffian quantum Hall state made simple---\\
multiple vacua and domain walls on a thin torus}

\author{E.J. Bergholtz}

\author{J. Kailasvuori}

\author{E. Wikberg}

\author{T.H. Hansson}

\author{A. Karlhede}

\affiliation{Department of Physics,
Stockholm University \\ AlbaNova University Center\\ SE-106 91 Stockholm,
Sweden}

\date{\today}

\begin{abstract}
We analyze the Moore-Read Pfaffian state on a thin torus. The known six-fold
degeneracy is realized by two inequivalent crystalline states with a four- and
two-fold degeneracy respectively.  The fundamental quasihole and 
quasiparticle excitations are domain walls between these vacua, and simple counting
arguments give a Hilbert space of dimension $2^{n-1}$ for $2n-k$ holes and $k$ particles at fixed positions
and assign each a charge $\pm e/4$.
This generalizes the known properties of the hole excitations
in the Pfaffian state as deduced using conformal field theory techniques. 
Numerical calculations using a model hamiltonian and  a small number of particles supports  
the presence of a stable phase with degenerate vacua and quarter charged domain walls also away 
from the thin torus limit. A spin chain hamiltonian encodes the 
degenerate vacua and the various domain walls.

\end{abstract}


\maketitle

One of the most intriguing aspects of the quantum Hall (QH) system is the
possibility of experimentally observing non-abelian statistics. In
particular it has been proposed that the fractional filling part of the
observed \cite{willett} $\nu = 5/2$ state is well described by the non-abelian Moore-Read, or
Pfaffian, wave function \cite{mr},
\be{mr} \Psi_{\textrm{Pf}} (\{ z_i\} ) =
\textrm{Pf}\left(\frac{1}{z_i-z_j}\right)  \Psi_{1/2} \ ,
\ee
where $ \Psi_{1/2} $
is the bosonic Laughlin state at $\nu=1/2$ and $z_i$ are the
complex electron coordinates in the plane. By now, a great deal has
been learnt about \pref{mr} and its quasihole excitations, and we
list some of the pertinent results: 
I. Eq.\pref{mr}  is the exact
ground state of a certain local three-body interaction \cite{greiter,rrhaldane}.
 II.
There are six degenerate ground states on a torus, and the
electronic wave functions are explicitly known \cite{greiter}. 
III. The quasiholes
have charge $e/4$ and  can only be created in pairs \cite{mr}. The
dimensionality of the Hilbert space for $2n$ holes at fixed
positions in the plane is $2^{n-1}$, and the wave functions in a particular
"preferred basis" have been constructed \cite{nayak,readrezayi}. 
IV.
 The quasihole wave functions are the conformal
blocks of a correlator in a $c=3/2$ rational conformal field theory
involving a bosonic vertex operator and a majorana fermion.
The conformal blocks have been explicitly constructed for four holes,
where the Hilbert space is two-dimensional. The braiding properties,
or monodromies, of the conformal blocks translate into non-abelian
statistics for the quasiholes \cite{mr,blokwen}. 
V. 
The ground state in eq.\pref {mr}
can be viewed as a triplet pairing state of composite
fermions, and the quasiholes as vortex excitations. The
pairing picture nicely explains the presence of quarter charged
holes \cite{greiter,readgreen}.

Finally we should mention that the great recent interest in the non-abelian QH
states has to a large extent been spurred by the proposals  to use them to
build decoherence free quantum computational devices \cite{Freedman}.

Recently, it was shown that studying the lowest Landau level (LLL) on a thin
torus, with circumference $L_1$,  both allows for a simple understanding of already established phenomena,
and for arriving at new results \cite{bk1,bk2}.  In particular, it was shown how the states
in the Jain series $\nu=p/(2pm+1)$ \cite{Jain} become gapped crystals, with a unit cell
of length $2pm+1$ (in units of the lattice spacing) and the fractionally charged excitations  appear as
domain walls between the $2pm+1$ different translational states of this
crystal. In the infinitely thin limit, also the $\nu = 1/2$ state
forms a crystal, which however melts at $L_1\sim 5.3$ (lengths are measured in
units of the magnetic length), and a gapless homogeneous state of neutral fermions forms. 
All these properties are consistent with known properties of
the bulk Laughlin and Jain states, and give a concrete realization of the dipole picture of
the gapless $\nu = 1/2$ state.
There is strong evidence, analytical and numerical, that all these states develop continuously 
into the bulk states as $L_1 \rightarrow \infty$ \cite{Haldane94,bk2,Lee05}.
In summary, we consider it established that
the main qualitative features of the bulk Laughlin/Jain states and the gapless
$\nu = 1/2$ state are present on the thin torus and that there is no phase transition 
as  the two-dimensional bulk case is approached.

In light of the above, we have studied the torus generalization of the Pfaffian
state \pref{mr} and its excitations on a thin torus. Despite recent progress on the 
construction of quasiparticle states \cite{hansson}, much more is known about quasiholes 
than of quasiparticles in the QHE. However, our construction is manifestly particle-hole symmetric 
and allows for a unified description of quasiholes and quasiparticles.
The analysis, to be given below, results in a simple and intuitive picture of
the degenerate ground states, and the quasiholes and quasiparticles as domain walls between them.
We obtain the general, $2^{n-1}$-fold degenerate, state with $k$ quasiholes and $2n-k$ quasiparticles. 
Using exact numerical diagonalization, we find that for a certain range of
pseudopotential parameters these quarter charged particles and holes are the lowest
energy excitations of systems with a small number of particles also at finite
$L_1$.

%
%
Defining the magnetic translation operators,
$t_1=e^{(L_1/N_s)\partial _x}, \, t_2=e^{(L_2/N_s)(\partial _y
+ix)}$, appropriate to the $A_y = 0$ (Landau) gauge, a basis of
lowest Landau level (LLL) single particle states  on a torus $(L_1,
L_2)$ is given by  $\psi_k=t_2^k \psi_0$, $k=0,1,...,N_s-1$, where
$\psi_0=\pi^{-1/4}L_1^{-1/2}\sum_n e^{inL_2x} e^{-(y+nL_2)^2/2}$.
Here $\psi_k$ is located along the line $y=-2\pi k/L_1$ and is a
$t_1$ eigenstate, $t_1\psi_k=e^{i2\pi k/N_s}\psi_k$. The quantum
number $k$ thus labels both the position in the $y$-direction and
the momentum in the $x$-direction. The many-body translation
operators $T_{\alpha}=\prod_{i=1}^{N_e} t_{i\alpha}$, ($t_{i\alpha}$
translates electron $i$) commute with a translationally invariant electron
interaction hamiltonian $H$.

A general $N_e-$particle state in the LLL is a linear
combination of states ${\rm det} (\psi_{k_1}(1)\ldots \psi_{k_{N_e}}(N_e))$, here
labeled by $n_0n_1\ldots n_{N_s-1} $, where $n_k=0,1$ and $\sum_{k=0}^{N_s-1}n_k=N_e$;
$T_2$ generates translations: $T_2n_0n_1\ldots n_{N_s-1} =n_{N_s-1}n_0\ldots n_{N_s-2} $.
As $L_1\rightarrow 0$ hopping becomes unimportant
and all energy eigenstates have the  charges frozen in a
regular lattice determined by the electrostatic interaction.
In a half-filled Landau level the ground state is
$101010...$---this is the obvious one-dimensional
limit when the electrons interact via a generic repulsive interaction. The interesting
question is now what 
happens when the length $L_1$ moves away from zero. In ref.
\onlinecite{bk1} it was shown that as hopping becomes more important, 
and for an unscreened Coulomb interaction at
$\nu = 1/2$, a gapless state obtained from the maximally hoppable state $01100110\dots 0110$
wins over the gapped crystal at $L_1\approx 5.3$, and it was later shown \cite{bk2} that the
resulting Luttinger liquid type state is well described by a Fermi sea of
composite fermions of the Rezayi-Read type \cite{rr}. From exact
diagonalization studies using an unscreened Coulomb potential, one also learns
that the gapless Rezayi-Read state is good at $\nu = 1/2$, while at $\nu = 5/2$ the
gapped Pfaffian state is favoured \cite{rrhaldane, morf}. The difference between the two cases is due
to the modifications in the short distance interaction caused by the different
one-particle states in the two Landau levels. 

On the torus, the Pfaffian state is six-fold degenerate rather than only two-fold as implied by the 
center of mass degeneracy. The technical reason for having the extra states is that
on the torus,
$
\Psi_{\textrm{Pf}} = \textrm{Pf}\left(\frac{1}{z_i-z_j}\right) \Psi_{1/2} \rightarrow \textrm{Pf}  \left(\frac{\vartheta_a(z_i-z_j)}{\vartheta_1 (z_i-z_j)}\right)\Psi_{1/2}^{(t)}
$,
where $\Psi_{1/2}^{(t)}$ is the torus version of $\Psi_{1/2}$, and $\vartheta_a(z)$ are Jacobi theta functions \cite{greiter}. 
The extra three-fold degeneracy 
corresponds to $a=2,3,4$.  Since the Pfaffian state is gapped
it is tempting to identify it with the crystalline state $A = 010101\dots 01$ in the thin limit, but
this, and its translated twin, $T_2 A$, only account for two of the six ground states.
Natural candidates for the other four are the four translations  $T_2^kB, k=0,1,2,3$ of the 
state $B = {\color{red} 01100110\dots 0110} $. We have explicitly verified that these are
the six ground states by projecting the Pfaffian state onto a single particle basis
and studying the thin limit: $a=2$ gives the two $A$ states, whereas $a=3,4$ give the four $B$ states.
Note that all the six ground states have the property that any four adjacent sites are populated by 
exactly two particles, and that they are the unique states with this property.\footnote{The Pfaffian states on the 
torus and the crystal states $A$ and $B$ have  the same
quantum numbers. The simultaneously conserved symmetries are generated by $T_1$ and $T_2^2$ and their 
eigenstates are $A$ and $B_{\pm}=B\pm T_2^2B$. It is easy to see that $B_{\pm}$ can be used in the construction 
of the domain walls instead of $B$ and $T_2^2B$.} 

In a state formed by joining different ground states  $ABAB\dots $, domain walls  with three and one electron on four adjacent sites, $AB\sim 1011$ and $BA\sim 0010$ respectively, are created,  
\be{AB}
 A  B  =  01010\ul{1}{\color{red} \ul{011}001100110011\ul0}\ul{010}1010101  \ \ \ .
 \ee
 (Note that because of the periodic boundary conditions both domain walls are present in the $AB$ state.)
It follows from the
Su-Schrieffer counting argument \cite{Schrieffer} that these domain walls have fractional charge $-e/4$ and $e/4$ respectively.\footnote{
Alternatively, notice that moving the primed electron in the configuration 
010\ul{1} {\color{red} \ul{011'}001100110} 
a single site gives 0101010\ul{1'}{\color{red} \ul{011}00110}, \ie the domain wall moves
four sites and thus carries charge $-e/4$. 
 }
Thus the state  $ABAB\dots B $ has an alternating sequence of 
positive and negative quarter charges---$AB$, in particular, contains one  quasiparticle-quasihole pair.

The four-fold degeneracy of $B$ compared to the two-fold degeneracy of $A$ leads to a degeneracy of the states with four or more excitations. Imagine 
inserting $B$-strings in a given  $A$ background. This can 
in general be done in two different ways  as illustrated by the following example,
\be{abab}
(ABAB)_1  &=&  010\ul{1}{\color{red} \ul{011}0011\ul0}\ul{010}101
                      0\ul{1}{\color{red} \ul{011}0011\ul0}\ul{010}101  \nonumber \\
          &\equiv& \!   01010\ul{1}{\color{red} \ul{011}0011\ul0}\ul{010}101
                      0\ul{1}{\color{red} \ul{011}0011\ul0}\ul{010}1  \nonumber \\    
(ABAB)_2  &=&  010\ul{1}{\color{red} \ul{011}0011\ul0}\ul{010}10101
                      0\ul{1}{\color{red} \ul{011}0011\ul0}\ul{010}1 \nonumber \\
\ee
where the $\equiv$ sign denotes equality up to a total translation. The single particle-hole state $AB$ in \pref{AB} is thus unique up to a translation, while the two states $(ABAB)_1$ and $(ABAB)_2$ cannot be translated into each other. Generalizing this we conclude that there are $2^{n-1}$ states of $n$ particle-hole pairs at fixed positions. 
In comparing the two states above, we notice that they differ only in that one of the $B$-segments is 
translated two lattice spacings. One might worry that this just corresponds to a shift of the positions of the domain walls, and would not imply the existence of many states at fixed positions. Note, however, that no  combination of rigid translations and {\em local} motion of the domain walls (where they stay separate) can transform the states into each other, and thus they belong to different topological sectors.\footnote{
Having an exact definition of the position of a domain wall is not necessary for these arguments,
but there is a rather natural prescription using the $B_\pm$ states introduced in  note 19.
With this definition the state counting for fixed domain wall positions agree with the one in the text which is however easier to visualize. }

\newcommand\bz {{\color{green}0}}
\newcommand\bo {{\color{green}1}}

To obtain general states with  quasiholes and/or quasiparticles one must insert extra empty sites and/or electrons. 
Define $A_0=A\bz=0101\dots 01\bz$ and $A_1=\bo A=\bo 0101\dots 01$. The state
$A_0$ has one extra 0 inserted---this excitation has, again by a straightforward counting argument, charge $e/2$. Similarly, $A_1$ has an excitation with charge $-e/2$. 
Joining these with $B$ one obtains the new domain walls $A_0B \sim 010\bz$, 
$BA_1 \sim 110\bo$ with charge $e/4$ and $-e/4$ respectively.
The domain walls $BA_0\sim BA \sim  0010 $ and $ A_1B \sim AB \sim 1011$ are the same 
as those already present in $AB$  \pref{AB}.  
Examples of a two-quasihole and of a two-quasiparticle state are,
\be{AB0}
 A_0  B  &=&  01 01\ul{01\bz}{\color{red} \ul{0}11001100110011\ul0}\ul{010}101   \nonumber    \\
  A_1  B & =& 1 01010\ul{1}{\color{red} \ul{011}0011001100\ul{110}\ul\bo}01010   \ \ \ .    
 \ee
A state with an arbitrary number of 
quarter charged holes and particles, in arbitrary positions, can be formed as  $X_1BX_2B\dots X_nB$, where $X_i \in \{A, A_0, A_1\}$.
Again disregarding a rigid translation, this state is $2^{n-1}$-fold degenerate for fixed positions of the particles and holes. 

In particular, the $2n$ quasihole states are $A_0BA_0B\dots B$ with degeneracy $2^{n-1}$ as for the $2n$ hole Pfaffian state on the plane. 
For eight and sixteen electrons, we have also explicitly verified that these states emerge as the leading terms in the  thin torus limit for the Pfaffian wave functions with two $e/4$ quasiholes (where $\vartheta_a (z_i-z_j) $ is replaced by
$\vartheta_a(z_i-z_j+{\tiny\frac{1}{2}} (\eta_1-\eta_2)  ) \vartheta_1 (z_i-\eta_1)\vartheta_1 (z_j-\eta_2) $ 
and the center of mass coordinate becomes $Z=\sum_i z_i + {\tiny\frac{1}{4}}(\eta_1+\eta_2)$ \cite{greiter}).

The six Pfaffian states are the exact ground states of a
hyperlocal three-body  interaction on the torus \cite{rrhaldane}---this holds
for general $L_1$ as it depends on the local properties only. 
The lattice hamiltonian \footnote{We give the matrix elements for the cylinder for simplicity.} takes the form $H_{3}=\sum_{\{k_i\}}
V_{\{k_i\}}c_{k_1}^\dagger c_{k_2}^\dagger c_{k_3}^\dagger c_{k_4}c_{k_5}c_{k_6}$ with
$V_{\{k_i\}}\propto\delta_{k_{123}, k_{456} }k_{12}k_{13}k_{23}k_{45}k_{46}k_{56}e^{-2\pi^2(\sum_i k_i^2-\frac 1 6 (\sum_i k_i)^2)/L_1^2}$, 
where $k_{ij}=k_i-k_j$ and $k_{ijk}=k_i+k_j+k_k$. 
In the thin torus limit, this implies  that the electrostatic energy is minimized by minimizing the number of sequences of four consecutive sites 
containing three electrons (or holes). The six states $A$ and $B$ above are the unique states at half-filling that have no such sequences.\footnote{$A$  actually has lower 
electrostatic energy than $B$---even as 
$L_1  \rightarrow 0$ there is a contribution from the hopping terms that makes the total energies for 
$A$ and $B$ equal. This is an artifact of the hyperlocal interation, for a more realistic longer-range interaction, hopping freezes out completely, leaving an entirely
electrostatic problem.} Such sequences of electrons (holes) are also absent from the states with quasiholes (quasiparticles).

\begin{figure}[h!]
\begin{center}
\resizebox{!}{70mm}{\includegraphics{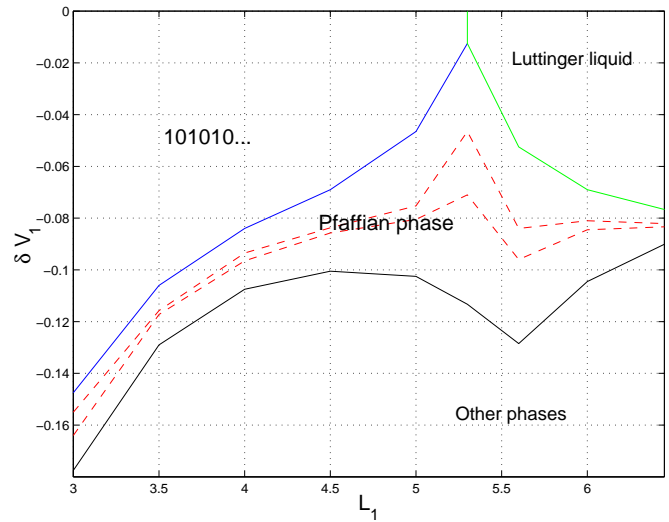}}
\end{center}
\caption{\textit{{\small  Phase diagram for a half-filled Landau level on the torus as a function $L_1$ and the pseudopotential parameter 
$\delta V_1$ ($\delta V_1=0$ corresponds to Coulomb interaction in the lowest LL). Results are obtained using exact diagonalization 
for eight electrons. The Pfaffian phase is found in the central part (solid lines) of the diagram. This is the region where the six 
lowest lying states are those that continuously approach the crystals $A$ and $B$. Within the dashed lines, these states are almost 
degenerate---they differ in energy by less than 10 {\%} of the gap to the next state. 
In the Pfaffian phase we observe quasiparticles and quasiholes of charge $\pm e/4$. In the top left the ground state is $101010...$ 
and the phase in the top right is the gapless Luttinger liquid phase described 
in ref. \onlinecite{bk1}. 
}
}}
\end{figure}

%
%
We have performed exact diagonalization studies of small systems that
corroborate the picture given above. Following Rezayi and Haldane
\cite{rrhaldane} we consider the electron gas on the torus as a
function of the pseudopotential parameter $\delta V_1$ and find that the
six Pfaffian states are favoured for a finite range in parameter space, see Fig. 1.
In particular, as the torus becomes thin, these states continuously
approach the crystalline states proposed above.
Exactly at half-filling we find that the low lying excited 
energy states are consistent with the creation of a single quasiparticle-quasihole pair. 
Due to their opposite charge they attract each other and the very lowest energies are 
obtained when they overlap. However, the entire low-energy spectrum is built up of states 
with different separations between the particle-hole pair.
Slightly away from half-filling we find ground states that have well-separated quasihole 
or quasiparticle excitations upon a Pfaffian background. In systems with two
quasiholes, the formation of $e/4$ charges is very clear and also
stable as $L_1$ increases from zero. As an example we find that
the ground state of the $\nu=8/17$ system with just eight particles evolves continuously
from ${\color{red} 0110011\ul{0}}\ul{010}101010$ (\ie a state with two $e/4$ quasiholes as far apart  as possible) 
into a charge density wave state with the same symmetry. This density wave is seen clearly 
throughout the Pfaffian phase in Fig. 1. Of course, 
we find---consistent with particle-hole symmetry---the same scenario in terms of $-e/4$ 
quasiparticles for $\nu=9/17$. Furthermore, 
we find that the lowest lying excitations of these systems are those that move the $\pm e/4$ charges closer together.

%
%
In ref. \onlinecite{bk1} it was shown that for $\nu = 1/2$, and a particular  choice of short-range hamiltonian relevant for  a thin torus the system  can be written as a  spin-half XY-chain by the mapping $10\rightarrow \uparrow$, $01\rightarrow \downarrow$. A spin flip, $\uparrow\downarrow\leftrightarrow\downarrow\uparrow$ corresponds to the nearest neighbour hopping $1001\leftrightarrow0110$ and the ground state emerges from the maximally hoppable state $B$.  By standard techniques the spin chain can be mapped onto free fermions, and it is natural to assume that a more general hamiltonian will correspond to a Luttinger liquid.

The most obvious way to generalize this description to the Pfaffian state is to consider the phase diagram for the spin-half model in the presence of anisotropic and more long range interactions. In addition to the gapless Luttinger liquid phase there are at least two gapped phases:  The Ising phase,
that will always win in the extreme thin torus limit, and a spin-Peierls phase. The latter is however not a candidate for the Pfaffian state; the spin paring breaks translational invariance, 
the ground state degeneracy on a torus in not six, and there are no quarter charged holes. It is an open question whether there is a QH counterpart to the spin-Peierls state.

The origin of these difficulties is that the above mapping of two sites to a single spin does
not allow for domain walls. To overcome this, we map each site to a spin such that the occupation number gives the $z$-component, 
$\sigma^z_i=2n_i-1$.
Remembering that the ground states are the states where any four adjacent sites have exactly two particles, suggests the hamiltonian 
$H_\textrm{p} = V \sum_i (\sigma^z_{i}+\sigma^z_{i+1}+\sigma^z_{i+2}+\sigma^z_{i+3})^2$, where $V>0$,
which clearly has the correct ground states. A  quarter charged quasiparticle (hole)  has one quadruple of sites with three electrons (holes), hence its excitation energy is $4V$.
This spin model is a frustrated antiferromagnetic spin chain, as can be seen by rewriting the hamiltonian as
$H_\textrm{p} =2V\sum_i (3\sigma^z_i\sigma^z_{i+1}+2\sigma^z_i\sigma^z_{i+2}+\sigma^z_i\sigma^z_{i+3})+\textrm{const.}$. 
The kinetic hamiltonian, $0110\leftrightarrow 1001$, is unfortunately somewhat complicated, $H_\textrm{k}=t\sum_i (\sigma_{i}^+\sigma_{i+1}^-\sigma_{i+2}^-\sigma_{i+3}^+ +\mathrm{h.c.})$. The hopping term lifts the degeneracy of the six ground states---this can however be compensated for by fine-tuning the $\sigma^z_i \sigma^z_j$ couplings.

In summary, we have presented a simple way to understand the vacuum degeneracy and the $\pm e/4$ charged quasiparticles and holes of the Pfaffian wave function in the thin torus limit. We have also given numerical evidence for this fractionalized phase to survive as the torus becomes thicker. In particular we found that the internal Hilbert space of a configuration of $2n$ particles and/or holes is $2^{n-1}$, and we should again stress that the quasiparticles enter in a natural way in our description. That the degeneracy of the internal quasihole Hilbert space agrees with the bulk state, strongly suggests that the non-abelian statistics also is present in the thin torus limit. Since the configuration space is one-dimensional and discrete, it is not clear how to define non-abelian statistics, but we might speculate that 
it would be encoded in properties of the (rather complicated) spin-chain defined above. 

While finishing this paper we became aware of that F.D.M. Haldane has obtained results similar to those presented here \cite{mars}. Shortly after this work, a closely related study of the bosonic Pfaffian state at $\nu=1$ appeared \cite{seidel}. 

This work was supported by the Swedish Research Council and by NordForsk.

\end{document}